\input harvmac

\overfullrule=0pt
\def\Title#1#2{\rightline{#1}\ifx\answ\bigans\nopagenumbers\pageno0\vskip1in
\else\pageno1\vskip.8in\fi \centerline{\titlefont #2}\vskip .5in}

scaled\magstep3

\font\ticp=cmcsc10
\font\secfont=cmcsc10

%
%
\baselineskip=18pt plus 2pt minus 2pt

\def\ajou#1&#2(#3){\ \sl#1\bf#2\rm(19#3)}
%
\def\CH{{\cal H}}


%


\def\s{\sigma}

\def\t{\tau}
\def\a{\alpha}
\def\b{\beta}
\def\d{\delta}

\def\n{\nu}
\def\p{\pi}

\def\G{\Gamma}

%

\def\[{\left [}
\def\]{\right ]}
\def\({\left (}
\def\){\right )}

\def\TrH#1{ {\raise -.5em
                      \hbox{$\buildrel {\textstyle  {\rm Tr } }\over
{\scriptscriptstyle \CH _ {#1}}$}~}}

\def\IZ{\relax\ifmmode\mathchoice
{\hbox{\cmss Z\kern-.4em Z}}{\hbox{\cmss Z\kern-.4em Z}}
{\lower.9pt\hbox{\cmsss Z\kern-.4em Z}}
{\lower1.2pt\hbox{\cmsss Z\kern-.4em Z}}\else{\cmss Z\kern-.4em Z}\fi}
\def\IC{\relax{\rm I \kern-.5em C}}
\def\IR{\relax{\rm I\kern-.18em R}}
\def\1{\relax 1 { \rm \kern-.35em I}}
\font\cmss=cmss10 \font\cmsss=cmss10 at 7pt

%

\def\frac#1#2{{#1 \over #2}}
\def\ie{{\it i.e.}}

\def\p+{{\partial_+}}

\def\half{{1 \over 2}}
\def\ket#1{|#1\rangle}

\def\Ab{\bar A}

\def\apm{\alpha^{\prime}}

%
\def\half{{1\over2}}

\def\kb{\overline k}
\def\qb{\overline q}
\def\lb{\overline l}
\def\Xb{\overline X}
\def\Nb{\overline N}

\def\ab{\overline a}
\def\zb{\overline z}
\def\Zb{\overline Z}

\def\nub{\overline \nu}
\def\taub{\overline \tau}

\def\rpart{Z \left[{\alpha \atop  \beta }\right] (\tau )}

\def\chartheta{\vartheta \left[ {\alpha \atop \beta }\right]}

\def\apm{\alpha^{\prime}}

\newbox\bigstrutbox
\setbox\bigstrutbox=\hbox{\vrule height12pt depth5.5pt width0pt}

\def\bigstrut{\relax\ifmmode\copy\bigstrutbox\else\unhcopy\bigstrutbox\fi}

\def\G{\Gamma}
\def\S{\Sigma}

\def\d{\displaystyle}

\def\sb{{\bar s}}

\Title{\vbox{\baselineskip12pt
\hbox{\ticp TIFR/TH/97-19}
\hbox{hep-th/9705239}
}}
{\vbox {\centerline{\bf NONRENORMALIZATION OF MASS}\vskip2pt
\centerline{\bf OF SOME NONSUPERSYMMETRIC STRING STATES}  }}
\centerline{{\ticp
Atish Dabholkar,  Gautam Mandal, and P. Ramadevi
}}
\vskip.1in
\centerline{\it Tata Institute of Fundamental  Research}
\centerline{\it Homi Bhabha Road, Mumbai, India 400005.}

\vskip .1in

\bigskip
\centerline{ABSTRACT}
\medskip

It is argued that the quantum correction to the mass of
some very massive, nonsupersymmetric states vanishes
 in inverse proportion to their tree-level mass to all
orders in string loops. This approximate
nonrenormalization can explain the agreement between
the perturbative degeneracy of these states and the Sen
entropy of the associated black holes.

\bigskip

\bigskip
\Date{}

\vfill\eject

\def\npb#1#2#3{{\sl Nucl. Phys.} {\bf B#1} (#2) #3}
\def\plb#1#2#3{{\sl Phys. Lett.} {\bf B#1} (#2) #3}
\def\prl#1#2#3{{\sl Phys. Rev. Lett. }{\bf #1} (#2) #3}
\def\prd#1#2#3{{\sl Phys. Rev. }{\bf D#1} (#2) #3}

\def\rmp#1#2#3{{\sl Rev. Mod. Phys. }{\bf #1} (#2) #3}
\def\cmp#1#2#3{{\sl Comm. Math. Phys. }{\bf #1} (#2) #3}

%

\lref\Mumford{D. Mumford, {Tata Lectures on Theta}, Birkh\"auser (1983).}

\lref\VeVe{E. Verlinde and H. Verlinde, \plb{192}{1987}{95}.}

\lref\Minahan{J. Minahan, \npb{333}{1990}{525} \semi
A. Tseytlin \plb{208}{1989}{221}.}

\lref\DhPh{E. D'Hoker and D. H. Phong, \rmp{60}{1988}{917}.}

\lref\DVV{R. Dijkgraaf, E. Verlinde and H. Verlinde,
\cmp{\bf 115}{1988}{649}.}

\lref\HoPo{G. Horowitz and J. Polchinski, hep-th/9612146.}

\lref\HaMo{J. Harvey and G. Moore, hep-th/9609017.}

\lref\GrMe{D. Gross and P. Mende, \plb{197}{1987}{129};
\npb{303}{1988}{407}. }

\lref\WiOl{E. Witten and D. Olive, \plb{78}{1978}{97}.}

\lref\DaHa{ A. Dabholkar and J. A. Harvey,
\prl{63}{1989}{478};
A. Dabholkar, G. Gibbons, J. A. Harvey, and F. R. Ruiz-Ruiz,
 \npb{340}{1990}{33}.}

\lref\DGHR{ A. Dabholkar, G. Gibbons, J. A. Harvey, and F. R. Ruiz-Ruiz,
\hfill\break \npb{340}{1990}{33}.}

\lref\DGHW{A. Dabholkar, J.~P.~Gauntlett, J.~A.~Harvey, and D.~Waldram,
\hfill\break \npb{474}{1996}{85}, hep-th/9511053.}

\lref\Dabh{A. Dabholkar, \plb{402}{1997}{53}; hep-th/9702050.}

\lref\Peet{A. W. Peet,  \npb{446}{1995}{211}, hep-th/9504097.}

\lref\Senentropy{A. Sen, {\sl Mod. Phys. Lett.} {\bf A10} (1995) 2081,
hep-th/9504147.}

\lref\Senblack{A. Sen, \npb{440}{1995}{421}, hep-th/9411187.}

\lref\Senfour{A. Sen, {\sl Int. J. Mod. Phys.}
{\bf A9}  (1994) 3707,  hep-th/9402002.}

\lref\LaLi{L. D. Landau and E. M. Lifshitz,
{\it The Classical Theory of Fields}, 4th edition, Pergamon Press (1989),
$\S{96}$.}

\lref\Duff{M. Duff, J. T. Liu, and J. Rahmfeld, \npb{494}
{1997}{161}, hep-th/9612015.}

\lref\DuRa{M. Duff and J. Rahmfeld, \npb{481}{1996}{332},
hep-th/9605085.}

\lref\KLMS{D. M. Kaplan, D. A. Lowe, J. M. Maldacena, and A. Strominger,
\prd{55}{1997}{4898}, hep-th/9609204.}

\lref\GHS{D. Garfinkle, G. Horowitz and A. Strominger,
\prd{43}{1991}{3140}.}

\lref\HoSt{G. Horowitz and A. Strominger, \prl{77}{1996}{2368},
hep-th/9602051.}

\lref\StVa{A. Strominger and C. Vafa, \plb{379}{1996}{99},
{hep-th/9601029}.}

\lref\CaMa{C. G. Callan and J. M. Maldacena,  \npb{475}{1996}{645},
hep-th/9602043.}

\lref\HLM{G. Horowitz, D. Lowe, and J. Maldacena, \prl{77}{1996}{430},
hep-th/9603195. }

\lref\DhMaWa{A. Dhar, G. Mandal and S.R. Wadia, \plb{388}{1996}{51},
hep-th/9605234.}

\lref\DaMa{S. Das and S. Mathur, \npb{478}{1996}{561},
hep-th/9606185; \npb{482}{1996}{153}, hep-th/9607149.}

\lref\MaSt{J. Maldacena and A. Strominger, \prd{55}{1997}{861},
hep-th/9609026.}

\lref\GuKl{S.S. Gubser and I. Klebanov, \npb{482}{1996}{173},
hep-th/9608108.}

\lref\CaGuKl{C. Callan, S.S. Gubser, I. Klebanov and A. Tseytlin,
\npb{489}{1997}{65}, hep-th/9610172.}

\lref\Mal{J. Maldacena, \prd{55}{1997}{7645}, hep-th/9611125;
S.F. Hassan and S.R. Wadia, \plb{402}{1997}{43},
hep-th/9703163.}

\lref\Das{S. Das, hep-th/9703146.}

\lref\analy{E. D'Hoker and D. H. Phong, \prl{70}{1993}{3292}.}

\lref\MoWe{J.L. Montag and W.I. Weisberger, \npb{363}{1991}{527}
.}

\lref\Moor{G. Moore, \plb{176}{1986}{369}. }

\lref\vvd{ R. Dijkgraaf, E. Verlinde and H. Verlinde, \cmp{115}{1988}
{649}. }

\lref\fay{ J.D. Fay, {\it Theta functions on Riemann surfaces},
Lecture Notes in Mathematics, Vol.352 (Springer, Berlin, 1973) .}
\newsec{Introduction}

Supersymmetric states, or `super states' for short, are the states that
belong to short representations of the supersymmetry algebra \WiOl. The
main significance of super states stems from the fact that the
semiclassical spectrum of these states is often not renormalized as a
consequence of supersymmetry. A super state preserves some of the
supersymmetries, and it follows from the supersymmetry algebra that it is
extremal, \ie\ its mass $M$ equals the absolute value of some charge $Z$.
The renormalization of its mass is therefore related to the
renormalization of the charge. With enough supersymmetries, the charge is
sometimes not renormalized, which then implies that the mass is also not
renormalized. Therefore, the states that are tagged by a particular mass
and charges at weak coupling continue to have the same mass and charges
even at strong coupling. Moreover, charge conservation together with
energetic considerations can usually ensure the stability of these
states.
One can thus learn much about the strongly coupled regime of a
theory from
the spectrum of super states computed at weak coupling.

It is interesting to know if there are any nonsupersymmetric states
for which such a nonrenormalization is possible.
Such states, if they exist, can provide important additional
information about the strongly coupled dynamics that does not rely
entirely on supersymmetry. In this paper we investigate
this question for a set of electrically charged, non supersymmetric
states
in toroidally compactified heterotic string theory. In this theory, which
has $N=4$ supersymmetry in four dimensions, the coupling constant, and
consequently the charges of these states are not renormalized. However,
the nonsuper states belong to long representations of the superalgebra,
so there is no exact relation any more between their mass and their
charges. Therefore, {\it a priori}, one expects nonzero and typically
large mass renormalization. These states are  not expected to be
stable either. In fact, for a very massive state, there is a large
phase space
because the state can decay into a large number of lower mass
states, and the
decay rate is expected to be large.
For the special states considered in this paper, we find, however, that
for a large enough tree-level mass, the mass-shift and the decay rate are
both vanishingly small to all orders in string loops.

To describe these states, let us consider, for definiteness, heterotic
string theory compactified on a six torus. The states are specified
by the
mass $ M$, the charge vector $(q, \qb )$ that lives on an
even, self-dual Lorentzian lattice $\G_{6, 22}$,
and the  right-moving and the left-moving
oscillator numbers $ N$ and $ \Nb$ respectively. For a general state, in
the NSR formalism, the Virasoro constraints can be written as
\eqn\massleft{
 M^2=\qb^2 + 2(\Nb - 1) = q^2 + 2(N - \half).
 }
Here and in
what follows, the quantities with the bar, like $ \qb$, are
left-moving, and
the corresponding unbarred quantities are right-moving; we have also
chosen $ \apm =2$. There are two special cases that are of interest:

$ \bullet$ $ N=\half$,  $ \Nb$ arbitrary.
These states are supersymmetric.
Their mass and charge are not renormalized \DaHa,
so the tree-level extremality relation, $ M=|q|$, is exact.

$ \bullet$ $\Nb =1$,  $N$ arbitrary. These states
are nonsupersymmetric, but classically extremal,
\ie, $ M= |\qb|$ at tree level.
Now, there is no supersymmetric nonrenormalization theorem
for the mass of these states.
So, in general,  the renormalized mass is not expected
to satisfy the tree-level extremality.

In this paper we shall  be interested in
the second set of states. One of our main motivation for
considering these states is their connection with
black hole entropy.
Let us  briefly explain why we choose these particular
nonsuper states  and why we expect  that their mass is not
renormalized.
Consider a set of states of given mass, charges,
and degeneracy  at weak  coupling.
As one increases the strength of the coupling,  the state
eventually undergoes a gravitational collapse to form a black hole.
Now, if for some reason the semiclassical spectrum of
these states is not renormalized, then  the degeneracy of states
is not expected to change
as we vary the coupling.
Therefore, the logarithm of the number of perturbative
states must agree with the Bekenstein-Hawking
entropy which counts the  corresponding black-hole
states. For the  supersymmetric states this
is certainly true.
As shown by Sen \Senentropy, the entropy given by the area
of  the stretched horizon of four dimensional, supersymmetric,
electrically charged,
black holes does agree, up to a proportionality constant,
with the statistical entropy given by the logarithm of the
perturbative degeneracy of the super states.
This works in other dimensions as well \Peet.
For the super states, this agreement is a consequence of
supersymmetric nonrenormalization of the spectrum.

What is more surprising is that a  similar agreement holds
\refs{\Duff, \DuRa}
also for
the set of nonsupersymmetric
but extremal states considered above.
Now, supersymmetry alone does not protect the spectrum from
getting  renormalized. On the other hand,
the agreement between the perturbative degeneracy
of states and  the black hole entropy would not be possible if
the quantum corrections to the tree-level mass were large.
We are thus led to the expectation that perhaps the spectrum of
these nonsuper states  does not get renormalized for
reasons other than supersymmetry.
Partial support for this expectation can be found by looking
at the classical self-energy of these states.
It was argued in \refs{\Dabh, \DGHW} that
the classical renormalization
computed from the self-energy of the fields around such a state
vanishes.  The perturbative quantum calculation that we describe
in this paper is in a complementary regime. The perturbative
calculation is performed in the flat space background;
it is valid in the regime when the coupling constant
is sufficiently small so that the gravitational field around the
state considered is small. {}From this point of view
the calculation of the classical field
energies is nonperturbative  because it takes
 into account the classical field condensate
formed around the state. Taken together, these two results
in the two regimes
do seem to support the expectation that the mass is not
renormalized, although we cannot rule out some other nonperturbative
corrections.

Even though one would like to prove the nonrenormalization
for all states with $ \Nb =1$ but arbitrary $ N$, we are able
to prove it only in the restricted regime $ 1\ll N\ll M^2$. More
precisely, the restriction on $N$ is that under $M \to \lambda M,
\, \lambda \to \infty$, we have $N \to N$.
So the states that we consider in this paper satisfy \massleft\
together with
\eqn\condition{\Nb = 1,\quad 1  \ll N \ll M^2.
}
It may appear that these states are `nearly supersymmetric,'
and therefore the mass renormalization should be small anyway.
However, it should be remembered that a nonsuper state
belongs to a long multiplet, whereas a super state belongs
to a short multiplet. So, there is no smooth limit that takes
one to the other. All that is expected from supersymmetry
is that when $ N=\half$
the mass renormalization should vanish.
So, the mass-shift is expected to be proportional to $ N-\half$, but, it
can depend,
in general, on any function of the mass $ M$. We find a
very specific dependence that the quantum correction to
the mass is bounded by a quantity inversely proportional to the
tree-level
mass. Thus, the  mass renormalization is vanishingly small even
for the state that is in a long multiplet
with  $ N$  much larger than $ \half$, as long as the tree-level
mass is sufficiently large.

This  paper is organized as follows.
In section 2, we give the form of a nonsupersymmetric, highly
massive vertex operator.  In section 3 we review the formal
definition of the  two point function  in an arbitrary genus
Riemann surface, mainly to set up the notations used
in the following sections. In section 4,
we concentrate on the one loop two-point function in detail.
As is well-known, formal string amplitudes are typically
divergent for on-shell external momenta and are well-defined
only after proper analytic continuation.
We describe the
analytic continuation required for the present two-point amplitude
in detail. Using the well-defined amplitude
obtained this way
we study its mass ($M$) dependence
and show that it is bounded above by a finite $M$
independent quantity. In section 5, we generalize
the one loop result to higher loops.
Finally, in section 6 we present the conclusions.

\newsec{Massive Vertex Operator}

In order to calculate the mass renormalization of the
states with $N = n + 1/2$, ${\bar N}=1$, we first
need to find their vertex operators. For simplicity we set
all Wilson lines to zero and choose
the right-moving and the
left-moving charges
 $ q^a, a=4, \ldots 9$ and $ \qb^{\ab},  \ab =4,...,9$
respectively to take values in a
lattice $\Gamma_{6, 6}$.
We write the right-moving momentum as
$k=(p,q)$ and the left-moving momentum as
$\kb=(p, {\bar q})$ where $ p^{\a}, \a =0,...,3$
is the momentum in the noncompact
four dimensions.
In the light cone gauge there are several possible states.
For example,  for spacetime bosons  we can have
\eqn\states{
{\bar \alpha}^{j}_{-1} \psi^{i_1}_{-\half}\ldots
\psi^{i_{2n+1}}_{-\half} \ket{k, \kb},\quad
{\bar \alpha}^{j}_{-1} \alpha^{i_1}_{-1}\psi^{i_2}_{-\half}\ldots
\psi^{i_{2n}}_{-\half} \ket{k, \kb},\quad
{\bar \alpha}^{j}_{-1}\alpha^{i_1}_{-1} \ldots \alpha^{i_n}_{-1}
\psi^{i_{n+1}}_{-\half} \ket{k, \kb},
}
and many other possibilities. To illustrate the main features of our
nonrenormalization theorem,
we shall describe the calculation
only for the  states
$ {\bar \alpha}^{j}_{-1}\alpha^{i_1}_{-1} \ldots \alpha^{i_n}_{-1}
\psi^{i_{n+1}}_{-\half} \ket{k, \kb}$ with totally symmetric polarization
for the right-movers.
We have also checked it for many
other states, but not for all possible states with $ N=n + \half$.
However, the main considerations
are sufficiently generic and we believe that the results are \
valid for all  states subject to the constraints \massleft\ and
\condition.

In covariant RNS formalism, in ghost number zero picture,
the vertex operator for this  specific state is given by
\eqn\vertex
{\eqalign{
 V(k, \zeta, \nu; \kb, \overline\zeta, \nub) =
&{1 \over  \sqrt{(-1)^{n+1}\, n!}}
{\bar \zeta_{A}} ({\bar {\partial}} {\bar {X}^{A}} )
\exp (i \bar k. \bar X)  \,\, \times\cr
 {}\times\zeta_{A_1 A_2 \ldots A_{n+1}}
&D^2 X^{A_1} D^2 X^{A_2} \ldots D^2 X^{A_n}D X^{A_{n+1}}
\exp (i k. X) \,\,\cr
}}
where  $D = \partial_{\theta} + \theta \partial_{\n}$ and
$X(\nu,  \theta)=X(\nu) + \theta \psi (\nu)$.
This is a physical state with  $ M=|\qb|$ provided
$\zeta_{A_1 A_2 \ldots A_{n+1}} $ is totally symmetric,
and
\eqn\vir
{\eqalign
{
{}& {\bar {\zeta}}.\bar k=0~,~
\zeta_{A_1 A_2 \ldots A_i \ldots A_j
\ldots A_{n+1}} \delta_{A_i}^{A_j}=0~,~
\zeta_{A_1 A_2 \ldots A_i \ldots A_{n+1}} k^{A_i}=0~.~\cr }
}

\newsec{The Two Point Function}
The mass-shift and the decay
rate of a perturbative string state can be extracted from the real
and the
imaginary part of the two point function of the corresponding vertex
operator. The two point function at order $ g$ is given by
\eqn\amp
{\eqalign{
   A(\zeta _i,k_i,\kb _i)=& \sum_{(\alpha ,\beta)} \int
           {d{\bf M}} d^2\!\nu _1 d^2\! \nu _2
                 \cdot \rpart \cdot \Zb (\taub)  \cr
\times &\langle V(k_1,\zeta _1,\nu _1;\kb _1,\overline\zeta
_1,\nub _1)
           \,V(k_2,\zeta _2,\nu _2;\kb _2,\overline\zeta _2,\nub _2)
                  \rangle  ,\cr }}
where $d{\bf M}$ is the  measure over
moduli space ${\cal M}_g$ of the genus $ g $ surface.
The details of the measure will not be important for
our discussion but can be found in \refs{\VeVe, \DhPh}.
$\rpart$ and $\Zb (\taub)$ are the partition functions
in the right and the left sector respectively
with $ (3g-3)$ insertions of
the $ b$ ghosts as well as $ (2g-2)$ insertions of picture-changing
operators  to soak up all the ghost zero modes for $ g\geq 2$
\refs{\VeVe, \DhPh}.
For genus one, the  volume of the
conformal killing vectors is factored out.
We have summed over the spin structures $(\alpha ,\beta )$.
$\langle V V \rangle$ denotes the correlation function
of the vertex operators for a given point in the
moduli space obtained by integrating over only the matter fields.

The various correlators on a Riemann surface of genus
$g$ can be readily evaluated.
The fermion propagator is given by the S\"zego kernel
\eqn\szego{  \langle \psi ^A (\nu _1) \psi ^B (\nu _2) \rangle
                         _{(\alpha ,\beta )} = {\eta^{AB}\over
E(\nu _{1},
\nu_2|\tau)} \,\,
{\chartheta (z|\tau )\over{\chartheta (0|\tau )}}\, ,  }
where $ \tau^{IJ}$ is the
period matrix, $E(\nu_1, \nu_2 |\tau)$
is the prime form, and $\chartheta (z |\tau )$is the usual
$\vartheta$ function with characteristics $ (\a, \b )$ \Mumford.
The vector $ z$ is defined by $z^I
\equiv \int_{\nu_1}^{\nu_2} \omega^I$ where
$\omega ^I(\nu), I=1,...,g$ are the holomorphic abelian differentials
on the genus $g$ surface. We note that
the $z^I$ coordinatizes the
Jacobian variety of the Riemann surface, $ \IC ^g / {(\IZ +\tau \IZ )}$,
so we can write $ z^I\equiv z^I_1 + i z^I_2 =\s^I_1 +\tau^{IJ}
\s^J_2$, such that
$ \s_1^I$ and $ \s_2^I$ take values over the unit
interval $ [0,1]$. Moreover,
$ \t^{IJ} = \t^{IJ}_1 +i\t^{IJ}_2$, $ z^{I}_2=\t^{IJ}_2 \s^J_2$.
The bosonic correlators can be read off from the correlators of
exponential insertions \refs{\DVV, \DhPh} :
\eqn\correlation{\eqalign{
    &\langle \prod_{i=1}^n
              {\rm e}^{ik_i\cdot X(\nu _i)}\,
                  {\rm e}^{i\kb _i\cdot\Xb (\nub _i)} \rangle
          =  {\delta(\sum k_i ) \delta(\sum   \kb_i )
          \over {Z^{\rm cl}}}\prod_{i<j} \, E(\nu _i, \nu_j|\tau
)^{k_i \cdot
k_j} \,
               {\overline E (\nub_{i}, \nub_j|\taub )}^{\kb _i
\cdot \kb _j}\,
                  \times \,\, \cr
     & \times   \sum_{(l, \lb )} {\rm exp}
              \biggl\{ i\pi \Bigl [ (l\cdot \tau \cdot l )
                   - (\lb \cdot\taub\cdot \lb) \Bigr ]
      + 2\pi i \Bigl [ (l\cdot \sum k_i \int ^{\nu
_i} \kern-6.0pt\omega )
               - (\lb \cdot \sum  \kb _i \int^{\nub
_i}\kern-6.0pt \overline\omega )
                    \Bigr ] \biggr\} \,\, ,\cr
}}
with
$ Z^{\rm cl} = \sum  {\rm exp}
\left\{ i\pi (l\cdot\tau\cdot l\, -\,
                 \lb \cdot \taub \cdot \lb )\right\} $.
Here
$l^A_I$ and $\lb^{\Ab}_I$, $ I=1, \ldots,g, A, {\bar A}=0, \ldots,
9$ are the
left and
right moving momenta
running through the loops: $\int_{b^I}\partial X^A =l^A_I$ and
$\int_{b^I}{\bar \partial}  {\bar X}^{\Ab} =\lb^{\Ab}_I$ where $
b_I$ are the $
b$-cycles
of the Riemann surface.  The sum over $ {\d{(l,\lb ) }}$ should be
understood
as an integral over the noncompact momenta $ l^\a_I, \a = 0, \ldots, 3$,
and a sum over
the compact momenta  $ (l^a_I ,\lb^{\ab}_I )$
which lie  on a  lattice $\Gamma_{6, 6}^g$:
\eqn\test{\sum_{\d{(l,\lb ) }} \equiv \int \prod_{\a, I}dl_I^{\a}
\sum _{\d{(l^a_I ,\lb^{\ab}_I )}} \, .}
To calculate  the correlators involving operators like  $
\zeta_A\partial X^A$,  we
calculate the correlators with $\exp (i \zeta_A  X^A )$, take the
derivative $ -i \partial$, and keep the term linear in $ \zeta$. To
simplify
this procedure it is convenient to write the polarization tensor
$ \zeta_{{A_1}\ldots {A_{n +1}}}$ formally as the symmetrized product
of vectors: $\zeta^1_{A_1} \zeta^2_{A_2}\ldots \zeta^{n+1}_{A_{n+1}}$.
Moreover, we use point splitting for composite operators.
For example,  we write
$ :\kern-.3em O_1 O_2(\sigma )\kern-.3em :$ as
$ O_{1}(\sigma_1 ) O_{2} (\sigma _2)$, calculate the correlators,
subtract the terms singular in $ (\sigma_1 -\sigma_2 )$ and then
take the limit $ \sigma_1\to \s$ and $ \sigma_2\to \s$.

We would like to emphasize that the two point function for
the nonsuper states is nonzero even after summing over spin structures.
In this respect they are very different from super states whose
two point function vanishes after summing over spin structure
as a consequence of supersymmetry. To see how it works in our case,
note that the two-point function  \amp\ contains
a term  $\langle \partial \psi \psi  (\n _1) \partial \psi \psi
(\n_2)\rangle$.
Using  expression
\szego, the fact that the spin-structure dependent part of the fermionic
partition
function is proportional to $\vartheta\kern1.0pt^{ 4}\kern-2.0pt
\left[ {\alpha
\atop \beta }\right](0|\t )$,  and the Riemann theta identity, we
see that
this term is nonzero  even after summing over spin-structures.

With these remarks we are now ready to evaluate the amplitude.
For simplicity we discuss the one-loop case first.

\newsec{One Loop}

The nonzero part of the two point function in this case can
be written as
\eqn\twopo
{\eqalign
{ A=& \int {d^2 \tau\over \tau^2_2}  d^2 z
 \, B({\bar \tau})
\sum_{(s, \sb )} C_{\left( s, \sb \right)} (\tau, \bar {\tau},
z, \bar z, n)
\, \sum_{(l, \lb)} D_s(\tau,  k,  z, l) {\bar D}_{\sb} ({\bar \tau}
,  \kb,
\zb, \lb)\, ,\cr}
}
where the sum is over  $s= 0,\ldots, n-1; \bar s=0,1$
and $B({\bar \tau})= {\bar \eta} ^{-24} ({\bar \tau})$.
The
quantities $C_{(s, \sb)}$ and $D_s$ are given by
\eqn\cfunc
{\eqalign{
C_{s, \sb}=& \,K_s\, {\bar K}_\sb \,
[\partial^2 \log E(z,\tau)]^{n-1-s} \,
[{\bar \partial}^2 \log {\bar E}
(\bar z, {\bar \tau})]^{1-\sb} \, [E(z, \tau)]^{-k^2} \,\cr
&[\bar E(\bar z, {\bar \tau}]^{- \bar k^2}\,
\exp[\pi(k^2 + {\bar k}^2 )\s_2 \cdot \tau_2 \cdot \s_2]
}}
where $$K_s= \{\zeta_{2s+1} \cdot \zeta_{2s+2}\} \ldots
\{\zeta_{2n+1} \cdot \zeta_{2n+2} \}, $$
$$ \bar K_0 = {\bar \zeta}_1 \cdot {\bar \zeta}_2 \qquad \qquad
\bar K_1 = 1 $$
and
\eqn\mass{D_s =
\left(  l \cdot \zeta^1 \right) \ldots (l \cdot \zeta^{2s} )
\exp \{i \pi (l \cdot\tau_1\cdot l +2 l\cdot k z_1 )\} \,
\exp\{-\pi (l + k \sigma_2)\cdot \tau_2 \cdot  (l + k \sigma_2)  \}~,
}
The equation for ${\bar D}_{\bar s}$ is similar.
We have carefully checked that
the delta-function contact terms coming from the contractions between
holomorphic and antiholomorphic operators vanish to give
this factorized form.

\subsec{ Analytic Continuation}


We  note that the two-point function, as
given by the integral \twopo, diverges
at the boundary of the  moduli space for on-shell
values of the momentum $k$. To obtain an well-defined
amplitude one needs to analytically continue the
external momenta to an unphysical region where all
modular integrals are convergent and then analytically
continue back to the physical region. Such an analytic continuation
relevant to the  four point superstring amplitude of
massless states at one loop has been done in
\refs{\analy,\MoWe}. We describe below the details
of the analytic continuation in the present case. As mentioned
in the introduction, we will later use the well-defined amplitude
obtained this way to study its $M$-dependence.

It is easy to see that  divergences of the above integral can
only come from the boundary of the moduli space ($ \tau_2 \rightarrow
\infty$ ). The integrand in eqn. \twopo\  at the large
$\tau_2$ limit is
\eqn\larget
{\eqalign{
 A_{\infty} =& \int {d^2 \tau \over \tau_2^2} \, d^2 z \,
\exp (+2 i \pi {\bar \tau})
\, [\exp \{ - i \pi z_1 ({\bar k}^2 -k^2) \}]\, \cr
{}&[\exp \{- \pi \s_2 \tau_2 (k^2 + {\bar k}^2) \}]
  \, \exp [\pi \tau_2 \sigma_2^2
(k^2 + \bar k^2) ] \, D_{n-1} {\bar D}_1 \, \cr
\cr}}
where
\eqn\rearr
{\eqalign
{
D_{n-1} {\bar D}_1 = & \sum_{l, \lb}
\left(  l \cdot \zeta^1  \ldots l \cdot \zeta^{2(n-1)} \right)\,
\left(  \bar l \cdot {\bar \zeta}^1  \,l \cdot {\bar \zeta}^2 \right )
\, \exp \{i \pi \tau_1 (l - \bar l^2)\} \, \cr
{}& \exp \{ 2 i \pi (\sigma_1 + \sigma_2 \tau_1)
(l \cdot k - \bar l \cdot \bar k)\}
\, \exp \{- 2 \pi \tau_2  \,(l+k \sigma_2)^2 \} \,
\exp \{ \pi \tau_2 (l^2 - \bar l^2) \} \,\cr
{}& \exp \{2 \pi \tau_2 \sigma_2 (l \cdot k - \bar l \cdot \bar k)\} \,
\exp \{ \pi \tau_2 \sigma_2^2 (k^2 - \bar k^2) \}\, . \,
\cr}}
Since the two point function \twopo\ is invariant under the
modular transformation $\tau_1 \rightarrow \tau_1 + 1$ for all $\tau_2$,
its assymptotic form as $\tau_2 \rightarrow \infty$ must also be
invariant under it. This implies
\eqn\modu
{\eqalign
{ +2 + (l^2 - {\bar l}^2) - \sigma_2 [ \,2n + 2 (l\cdot k -
\bar l \cdot \bar k)\, ]= 2 r \, }
}
where $r$ is an integer.  Furthermore,
since this relation should be valid for arbitrary $\s_2$,
the term in square bracket has to be be zero\foot{%
The same conclusion can be arrived at if one performs
the $\s_1$ integration first.}.
Now, doing the $\tau_1$ integration in \larget,
we get
\eqn\diver
{\eqalign{
A_{\infty}=&\int {d \tau_2 \over \tau_2^2} \,d^2 z  \,
\exp \{\,2 \pi \,\tau_2 \,\sigma_2 \,(1- \sigma_2) (-k^2) \, \}
\sum_{l, \bar l}\, \delta_{(l^2 - \bar l^2), -2}\,\cr
{}&\delta_{(l\cdot k - \bar l \cdot \bar k), n}\,
\left(  l \cdot \zeta^1  \ldots l \cdot \zeta^{2(n-1)} \right)\,
\left(  \bar l \cdot {\bar \zeta}^1  \,l \cdot {\bar \zeta}^2 \right )
\exp \{\,- 2 \pi \tau_2 (l + k \sigma_2)^2\, .\}
\cr}}
Clearly, the integral is divergent if we impose the on-shell
condition on the integrand. We define the above integral by
analytically continuing in the  complex plane of $s = - p_\mu p^\mu$.
The integral is then convergent in the region
\eqn\domain
{\eqalign{
{\rm Re} (-k^2)=& {\rm Re} (-p^2 - q^2) < 0 \,}}
The result can be reevaluated at the on-shell values of the
momenta and it is well-defined.


\subsec{Bound on the two point function}


It is easy to see from \twopo\ that the quantities $ C_{(s, \sb)}$
do not depend on $ M$, but  only on
$ n$,  so in the limit
$ n\ll M^2$ considered in  \condition,  any
mass dependence, if at all,
comes from the functions $ D_s$ and $ {\bar D}_{\sb}$.
We now show  that the amplitude \twopo\
is actually bounded above by an $M$-independent quantity.

It is  convenient to divide $A$ into two parts:
\eqn\tacy {
A= A_1 + A_2
}
where $A_1$  and $A_2$ correspond to replacing  $B({\bar \tau})$
in \twopo\  with $\exp (2 i \pi {\bar \tau} )$ and
$ \{ \,({\bar \eta} ({\bar \tau})\,)^{-24} - \exp ( 2 i \pi
{\bar \tau}  )\, \}$ respectively. $A_1$ contains the
(unphysical) tachyonic divergence, which is
removed by the integration over $\tau_1$.
$A_2$ does not contain any such divergence
and the modulus of the integrand in  $A_2$ is
itself finite over the  entire  moduli space.

Now, clearly
\eqn\twoprime
{ |A| \leq  |A_1| + |A_2|  }

We will first show that $|A_2|$ is bounded above by a
finite $M$-independent quantity. Using the fact that the absolute
value of a sum is less than the sum of the absolute values, we have
\eqn\twoprime
{\eqalign
{|A_2| \leq  A' \equiv & \int {d^2\tau\over \tau_2^2}
d^2 z
 \, |B({\bar \tau})| \sum_{(s, \sb )}
|C_{( n-1-s, 1-\sb )} (\tau, \bar {\tau}, z, \bar z, n)|
\, \sum_{(l, \lb)} |D_s(\tau,  k, z, l)|\, |{\bar {D}}_{\sb} ({\bar
\tau}, \kb,
\zb, \lb)| }
}
Now, note that, because
$ \tau_2$ is a positive definite,
$A'$ is finite for all values of  $\tau_2$.
Furthermore, it depends on the external charges $ (k, \kb )$
only through the quantities $  (k \s_2^I, \bar k \s_2^I) $
which can always be  brought inside the unit cell  by a lattice
shift.
In other words, $(k\sigma_2, \bar k \sigma_2) =
(L, \bar L) + (\delta k, \delta \bar k )$,
where $ (L, \bar L)$ is a lattice vector and $ (\delta k, \delta \bar k)$
 lies inside the unit cell of the lattice.
 The summation variables $(l, \bar l)$
can be suitably shifted to replace $(k \sigma_2, \bar k \sigma_2)$
by $(\delta k, \delta \bar k )$ in the exponents of $|D\bar D|$
in \twoprime. The
extra terms
involving $L\cdot\zeta$ and $\bar L\cdot\bar \zeta$ that appear
in \twoprime\
in the process depend on $(k, \bar k )$. However,
since $k\cdot\zeta =0$
we have the condition that $ L\cdot \zeta = -\delta k \cdot \zeta$
(and similarly for the barred quantities). As a result, the
entire quantity $A'$ depends only on $(\delta k, \delta
\bar k)$.
Now, when $M \to \lambda M, \lambda \to \infty$, some of
the components of the charges $(k, \bar k )$ must go to
infinity  to satisfy the conditions \massleft\ and
\condition. However, $(\delta k, \delta \bar k)$ varies
inside a bounded domain (the unit cell) whose limits are
independent of $M$,
and  $A'$ is finite for all $(\delta k, \delta \bar k)$.
Therefore,  we conclude that $ A'$  and, in turn $ |A_2|,$ is
bounded above by a finite quantity independent of $M$.

Now, we are left with the task of showing that $|A_1|$ is
also bounded by an $M$ independent finite quantity.
For $A_1$, we will perform the  $\tau_1$
integration so that the unphysical tachyonic divergence
is removed.
Let us first consider $A_{1\infty}$
which includes only the leading
term in the large $\tau_2$ expansion of
$C_{s,\sb}$ (eqn. \cfunc). Doing the  $\tau_1$ integration
gives eqn. \diver. Now, since the
first exponential, with the analytically continued $k$ defined
is less than one, we have
\eqn\largat
{\eqalign{
A_{1\infty} \leq & \int {d \tau_2 \over \tau_2^2} \,d ^2 z\,
\sum_{l, \bar l}\, \delta_{(l^2 - \bar l^2), -2}\,
\delta_{(l\cdot k - \bar l \cdot \bar k), n}\cr
{}&\,
\left(  l \cdot \zeta^1  \ldots l \cdot \zeta^{2(n-1)} \right)\,
\left(  \bar l \cdot {\bar \zeta}^1  \,\bar l \cdot {\bar \zeta}^2
\right )
\exp \{\,- 2 \pi \tau_2 (l + k \sigma_2)^2\, \}
\cr}}

The Kroenecker delta's restrict the lattice sum to a
subset, but since the terms appearing in the sum
are all positive, the restricted sum is less than or
equal to the unresrticted sum, that is,
\eqn\largebt
{\eqalign
{A_{1 \infty} \leq & \int {d \tau_2 \over \tau_2^2} d ^2 z\,
\sum_{l, \bar l} \,
\left ( l\cdot \zeta^1 \ldots l \cdot \zeta^{2(n-1)} \right) \,
\left (\bar l \cdot {\bar \zeta}^1 \bar l \cdot {\bar \zeta}^2 \right)
\,  \exp \{-  2 \pi \tau_2 (l + k \sigma_2)^2 \}\, .
}}
Just as in the case of $A'$ \twoprime, we can now shift the
summation variable $(l, \bar l)$ of the lattice to
show that \largebt\ is bounded
above by a finite $M$ independent quantity.

The above arguments can also be made for contributions
to $A_1$ coming from the subleading terms in the large $\tau_2$
expansion of $C_{s, \bar s}$. One can again show that
the modulus of each
of these terms is bounded above by a finite $M$-independent
quantity. Since there are an infinite number of
terms, one of course needs to show that the series converges.
This is easy to show using the fact
the higher order terms in the large-$\tau_2$
expansion are exponentially damped.

To summarize, we have first described the analytic continuation in
the non-compact momentum plane to make the
two point amplitude well-defined. Then, we seperated the amplitude
into two terms mainly to handle the unphysical
tachyonic divergence and showed that the modulus of each of
the terms, and hence
the modulus of the one-loop two-point function
$|A|$, is bounded above by a finite $M$-independent quantity.

\newsec{Generalization to Higher Loops}

We  now extend the result of the previous section to higher
genus amplitudes. The two-point function is given by
\eqn\twopohigher
{\eqalign
{ A=& \int d{\bf M}  d^2 \nu_{1} d^2 \nu_{2}
 \, B({\bar \tau}) f(\tau, \nu_1, \nu_2)
\sum_{(s, \sb )} C_{\left( s, \sb \right)} (\tau, \bar {\tau},
\nu_1, \nu_2, {\bar \nu}_1, {\bar \nu}_2, n)\, \cr
\,& \sum_{(l, \lb)} D_s(\tau,  k,  \nu_1, \nu_2, l)
{\bar D}_{\sb} ({\bar \tau}
,  \kb, {\bar \nu}_1, {\bar \nu}_2, \lb)\, ,\cr}
}
where $C$, $D$ and  $\bar D$ are given by:
\eqn\chfunc
{\eqalign{
C_{s, \sb}=& K_s\, \bar K_\sb \,[\partial_{\nu_1} \partial_{\nu_2}
\log E(\nu_1,\nu_2 \vert\tau)]^{n-1-s} \,
[{\bar \partial}_{{\bar \nu}_1} {\bar \partial}_{{\bar \nu}_2} \log
{\bar E}
({\bar \nu}_1, {\bar \nu}_2 \vert {\bar \tau})]^{1-\sb} \, \cr
\, &[E(\nu_1,
\nu_2 \vert \tau)]^{-k^2} \,
[\bar E({\bar \nu}_1, {\bar \nu}_2 \vert {\bar \tau}]^{- \bar k^2}\,
\exp[\pi(k^2 + {\bar k}^2 )\s_2 \cdot \tau_2 \cdot \s_2]
}}
with $K_s, \bar K_\sb$ being the contraction of the polarization tensors
as defined in the one loop case
and
\eqn\massa
{\eqalign{D_s =&
\{  l \cdot \zeta^1 \omega(\nu_1)\,\}
\,\{  l \cdot \zeta^2 \omega(\nu_2)\,\}
\ldots \{l \cdot \zeta^{2s-1} \omega(\nu_1)\,\}
\{l \cdot \zeta^{2s} \omega(\nu_2)\,\} \cr
&\exp \{i \pi (l \cdot\tau_1\cdot l +2 l\cdot k z_1 )\} \,
\exp\{-\pi (l + k \sigma_2)\cdot \tau_2 \cdot  (l + k \sigma_2)  \}~.
}}
The equation for ${\bar D}_{\bar s}$ is similar.
$B(\bar \tau) = {\rm det}^{-12} (- \nabla^2)$ is the genus
$g$ determinant of 24 left-moving bosons. The function
$f(\tau, \nu_1, \nu_2)$ is the product of the right-moving
bosonic determinants and the spin-structure-summed
two-point function  $\langle \partial_{\nu_1}\psi \psi \,
\partial_{\nu_2} \psi \psi \rangle$. For our purpose
we need to know only the asymptotic properties of  $B$ and
 $f$  at the boundary of the
moduli space which we will mention shortly.

\subsec{Analytic Continuation}

As in the one-loop case, the integral expression \twopohigher\
is naively divergent at the boundary of the moduli space and
we need to define the amplitude through
analytic continuation. For consistency, the analytic
continuation employed in eq. \domain\ should work here too.
To see that it does, we need to consider in somewhat detail the boundary
of the moduli space of the higher genus Riemann surfaces. This
is a rather well-studied subject \refs{\Moor, \vvd, \fay}. For our
purposes, we can regard the boundary of the moduli space as
consisting of two distinct kind of degenerations of the
Riemann surface:

1. {\sl Pinching along a  homologically trivial cycle}:


Under this, a genus $g_1+g_2$  Riemann surface splits
into two disjoint Riemann surfaces (having
one puncture each)  of genuses $g_1$ and $g_2$.
It is not difficult to see that such a pinching factorizes
the present two-point function in such a way that one of the
factors always equals one-point function of a tachyon at
some loop. Since the latter quantity is zero, the contribution
of this kind of degeneration is zero. We shall therefore
consider degeneration along only the homologically nontrivial cycles
henceforth.

2. {\sl Pinching along a  homologically nontrivial cycle}:

Under this kind of a degeneration a genus $g+1$ surface
reduces to a genus $g$ surface with two punctures.
We need to know how the various factors in \twopohigher\ behave
under this kind of a degeneration. The behaviour of $C, D, \bar D$
can be read off from the behaviour of the period matrix $\tau$
and the abelian differentials $\omega^I$. The latter are
given as follows. Let $\tau^{IJ}, I, J=1,2, \ldots, g$ be the
period matrix and $\omega^I, I=1,2,\ldots, g$ the abelian
differentials of the genus $g$ Riemann surface obtained
after the degeneration. Also, let $a$ and $b$ be the two
punctures. Then near the boundary of the moduli space corresponding
to this
degeneration, the period matrix and abelian differentials
of the original genus $g+1$ Riemann surface is given by
\eqn\dvv{\eqalign{
\tau \rightarrow &\left [\matrix{
\tau_{I J} & \int_a^b \omega_I\cr
\int_a^b \omega_J &  \tau_{00}  \cr
}\right ] ,\cr
\omega_0(\nu_1) \rightarrow &\partial_{\nu_1}
 \log\left[ { E(a,\nu_1)\over E(b, \nu_1)} \right].\cr
}}
where $$\tau_{00} \rightarrow {-i \over \pi}
( { \log t \over 2} -  \log E(a,b))\, + O(t) \, \, \, {\rm as}
\, \, t \rightarrow 0.$$
Keeping in mind that there is an unphysical tachyon
in the ${\bar L}_0 =1$ spectrum of the leftmoving
bosonic string (prior to the Re $\tau$ integration),
we can infer the following asymptotic form of $B(\bar \tau)$:
\eqn\basym
{B(\bar \tau)= (\bar q_0)^{-2} B_1(\bar q_0, \bar q_I) +
B_2 (\bar q_0, \bar q_I)}
where $B_2$ is finite over the entire moduli space and
$B_1$ is finite at the specific degeneration of the
Riemann surface under consideration. Since there is no such
divergence from the right sector, the asymptotic form
of $f$ is:
\eqn\fasym
{f(\tau, \nu_1, \nu_2) = [1 + O(q_0^2)] f_1 (q_I, \nu_1, \nu_2)}
which is finite over the entire moduli space.
Here $q_I = \exp[i \pi \tau_{II}]$ and similarly for
the barred terms.

Combining all this, the asymptotic behaviour of $A$ at
this degeneration is given by
\eqn\twopoasym
{\eqalign
{A_\infty = &\int d{\cal M} d^2 \tau_{00}\int
   d^2 \nu_1 d^2 \nu_2 (\bar q_0)^{-2}
 [\exp \{ - i \pi z_1^0 ({\bar k}^2 -k^2) \}]\, \cr
{}&[\exp \{- \pi \s_2^0\, {\rm Im}\tau_{00}\, (k^2 + {\bar k}^2) \}]
  \, \exp [\pi \s_2^0\, {\rm Im}\tau_{00}\, \sigma_2^0
(k^2 + \bar k^2) ] \, D_{n-1} {\bar D}_1 \, \cr
}}
where $d{\cal M}$ is a measure factor whose explicit form is not
important for our purpose except  that it is independent
of   ${\rm Re} \,\tau_{00}$. We have explicitly verified this
fact in the two-loop case \Moor.
For higher loops the measure factors in terms
of the period matrix  are quite complicated for explicit
verification. However,  from the form of the
measure in the  Fenchel Nielsen parametrization \DhPh\
we believe that it is true also at higher loops.

Going through steps similar to the ones in the one-loop case
(i.e imposing modular invariance ${\rm Re} \,\tau_{00} \rightarrow
{\rm Re} \,\tau_{00} +1$ and then doing the ${\rm Re}\,\tau_{00}$
integration), we arrive at
\eqn\something{\eqalign{
A_\infty = &\int d{\cal M} d {\rm Im} \tau_{00}
 \int d^2 \nu_1 d^2 \nu_2
 \exp\{2 \pi \,{\rm Im}\tau_{00}\, \s_2^0
(1-\s_2^0)(-k^2)\}
\sum_{l, \bar l}\, \delta_{(l_0^2 - \bar l_0^2), -2}\,
\cr
&\delta_{(l_0\cdot k - \bar l_0 \cdot \bar k), n}
\{  l \cdot \zeta^1 \omega(\nu_1)\,\}
\,\{  l \cdot \zeta^2 \omega(\nu_2)\,\}
\ldots \{l \cdot \zeta^{2s-1} \omega(\nu_1)\,\}
\{l \cdot \zeta^{2s} \omega(\nu_2)\,\}
\cr &
\{  \bar l \cdot {\bar \zeta}^1 \omega({\bar \nu}_1)\,\}
\,\{ \bar l \cdot {\bar \zeta}^2 \omega({\bar \nu}_2)\,\}
\, \ldots \,
\{  \bar l \cdot {\bar \zeta}^{\bar s -1} \omega({\bar \nu}_1)\,\}
\,\{ \bar l \cdot {\bar \zeta}^{\bar s} \omega({\bar \nu}_2)\,\} \cr
&\exp\{-2\pi {\rm Im} \tau_{00}
 (l^0 + k \sigma_2^0)^2  \}~\, {\cal L}_g {\bar {\cal L}}_g
}}
where
\eqn\latti
{\eqalign
{{\cal L}_g=&
\exp \{i \pi (l^I \,{\rm Re} \tau_{IJ} \, l^J +2 l^I\cdot k z_1^I )\} \,
\exp\{-\pi (l + k \sigma_2)^I \cdot {\rm Im} \tau_{IJ}
\cdot  (l + k \sigma_2)^J  \}~.
}
}
and its conjugate is ${\bar {\cal L}}_g$.

We see, therefore, that the amplitude is well-defined under the
same analytic continuation \domain\ as used in the one-loop case.

\bigskip

\subsec{Bound on the two-point function}

We  now  prove $M$-independence of the two-point function
 in higher loops.

Consider the moduli
space after pinching any one non-zero homology
cycle labeled by `0',
As in the one-loop case, we  substitute \basym\ for $ B(\bar\tau)$
in \twopohigher,
and  divide $A$ into two parts:
\eqn\tacyhigher {
A= A_1 + A_2,}
Here  $ A_1$ and $ A_2$ correspond to the two terms
 $(\bar q_0)^{-2} B_1(\bar q_0, \bar q_I)$
and $B_2(q_0,q_I)$ in \basym\ respectively.

Since $B_2(\bar q_0, \bar q_I)$ is finite,
the integrand of $A_2$ is finite over the entire
moduli space. We can show that the modulus of $A_2$
is bounded above by a finite $M$ independent quantity
following the steps presented for one-loop.

$A_1$ contains the unphysical tachyonic divergences.
When only cycle `0' is pinched and no other
then we can deal with the tachyonic divergence as in the one-loop case.
At higher loops, however,  we have to take
into account simultaneous pinching of more than one
non-trivial homology cycles. For example, if an additional
cycle labeled by `1' is pinched along with `0' then we
have additional divergences.
In this case $ B_1(\bar\tau)$ can be written as
\eqn\bonetau{
(\bar q_0)^{-2} B_1(\bar\tau) = (\bar q_0)^{-2} B_1^{(1)}(\bar q_0, \bar q_I) +
(\bar q_1)^{-2} B_1^{(2)}(\bar q_0, \bar q_I)
+ (\bar q_0)^{-2} (\bar q_1)^{-2} B_1^{(3)}(\bar q_0, \bar q_I) ,}
where the $B$ coefficients on the right have no singularities
at either of the two degenerations.
Correspondingly $ A_1$ now has three possible
divergent terms
\eqn\split
{A_{1} = A_1^{(1)} + A_1^{(2)} + A_1^{(3)}.}

It is now evident that $A_1^{(1)}$ and $A_1^{(2)}$
are similar to the one loop $A_1$  which can
be handled by integrating the real part
of the appropriate period matrix element.
So, these two terms are again bounded by some
finite $M$ independent quantity.

The non-trivial term different from the one loop is
$A_1^{(3)}$. We have to integrate over the real part
of both the period matrix elements which dominate
in the simultaneous pinching of the two
homologically non-zero cycles.
This will lead to a restricted lattice sum
which can again be handled in exactly the same
fashion as for the one loop.

It is straightforward to extend the procedure of
seperating $A_1$ \split\ into various terms under simultaneous
pinching of arbitrary number of non-zero
homology cycles at every order in string loop and
showing that each of the terms is finite and bounded above
by an $M$ independent finite quantity.

\newsec{Conclusions}

The real part of
$ A$ equals $ \delta M^2$ and the imaginary part equals,
by the optical theorem,  $ \Gamma M$,
where $ \Gamma$ is the decay rate.
Since $|A|$ is bounded by an M independent
constant, it follows that both the mass correction
$ \delta M $ as well as the width  $ \Gamma $ are vanishingly small
if the tree-level mass is sufficiently large:
\eqn \inequality {\delta M \le O(1/M) \, ; \quad\, \Gamma \le O(1/M).}

One immediate consequence of our result is that it partially answers
our original question of why the degeneracy of even the nonsuper
states agrees with the Sen entropy of the associated black holes.  We
have proved here a perturbative nonrenormalization valid to all loops.
The agreement with the entropy suggests that it should be valid even
nonperturbatively.  There are a number of related issues that are
currently under investigation. Besides the entropy, the gyromagnetic
ratios of the nonsuper states are also known to be in agreement with
those of the associated black holes \refs{\DuRa, \Duff}. We therefore
expect a similar nonrenormalization of the gyromagnetic ratio as well.
The perturbative heterotic state considered here, which has winding
and momentum along the internal torus is dual to a D-string in Type-I
theory. It is interesting to know if a similar nonrenormalization
holds for the D-string. In particular, the large mass limit considered
here appears to be related to the large-N limit in the gauge theory on
the D-string.  More generally, one expects a similar
nonrenormalization for a host of magnetically charged as well as
dyonic states that are S-dual to the purely electrically charged
states discussed here.

In this paper we have discussed black holes with zero area.  For black
holes with nonzero area, both in the supersymmetric as well as nearly
supersymmetric cases, there is already an impressive agreement between
the degeneracy of the string states and the Bekenstein-Hawking entropy
\refs{\StVa, \HoSt, \CaMa}.
Furthermore, the emission and absorption properties of these black holes
also match those of the corresponding string states
\refs{\DhMaWa, \DaMa, \MaSt, \GuKl, \CaGuKl}.
Indeed, the entropy-matching seems to work even for a few nonsuper states
\refs{\Dabh, \HLM, \KLMS}.
It would be interesting to see if  these states satisfy
mass-nonrenormalization
of the kind presented here.
Previous discussions
of nonrenormalization in the context of nearly supersymmetric black holes
with nonzero area can be found in \refs{\Mal, \Das}.

We end with a few comments and speculations.  Our results are in
accord with the general correspondence, pointed out in \HoPo, between
perturbative string states and black holes.  By this correspondence
the two degeneracies match at a specific value of the coupling where
the quantum correction to the spectrum is becoming appreciable.
However, if there is no renormalization of the spectrum, as in our
case, the degeneracies can be compared at all values of the coupling.
The nonrenormalization that we have found is quite surprising because
it is not a consequence of any apparent symmetry. It may be that there
is a hidden gauge symmetry of string theory that is responsible for
this nonrenormalization.

Our results indicate that there is an infinite tower of very massive
nonsuper states with $ \Nb = 1$ and arbitrary $N$, which is very
similar to the infinite tower of super states with $ N=\half$ and
arbitrary $ \Nb$. The spectrum of super states has important
application to the dynamics of the theory.  For instance, in $ N=2$
string theories one can construct a generalized Kac-Moody Lie
superalgebra in terms of super states which governs the form of the
perturbative superpotential \HaMo. It would be interesting to see if
the nonsuper states discussed here or their duals can be used to
obtain additional insight into the dynamics.

\bigskip
\leftline{ \secfont Acknowledgements}
\bigskip

We would like to thank Indranil Biswas, Pablo Gastesi,
Ashoke Sen, and Lenny Susskind for useful
discussions, and the organizers of the 1996 Puri Workshop for inviting
us to a very stimulating meeting where this work was
initiated. G.M. would like to acknowledge the hospitality of CERN
theory division where part of the work was done.
\vfill
\bigskip

\listrefs
\end